\documentclass[a4paper,11pt]{article}
\usepackage{pos}

\newcommand{\qedl}{\mathrm{QED}_{\mathrm{L}}}

\newcommand{\qedr}{\mathrm{QED}_{\mathrm{r}}}

\newcommand{\qedc}{\mathrm{QED}_{\mathrm{C}}}
\newcommand{\qedm}{\mathrm{QED}_{\mathrm{M}}}

\title{Structure-dependent electromagnetic finite-volume effects to the hadronic vacuum polarisation}
\ShortTitle{Structure-dependent electromagnetic finite-volume effects}

\author*[a]{Nils Hermansson-Truedsson}

\affiliation[a]{The Higgs Centre for Theoretical Physics, School of Physics and Astronomy, The University of Edinburgh\\
  Mayfield Rd, Edinburgh, EH9 3JZ, United Kingdom}

\emailAdd{nils.hermansson-truedsson@ed.ac.uk}

\abstract{In this talk we present some preliminary results and discuss the prospects of determining the leading structure-dependent finite-volume effects in the hadronic vacuum polarisation associated to order $e^2$ electromagnetic corrections. In the quantum electrodynamics prescription $\qedl$ these arise at order $1/L^3$ in the large-volume expansion, which is also the leading order because of the neutrality of the currents defining the underlying correlation function. Knowing the size of the finite-volume effects in question is relevant for determinations of the leading isospin-breaking corrections to the muon anomalous magnetic moment coming from the hadronic vacuum polarisation. 
}

\FullConference{The 41st International Symposium on Lattice Field Theory (Lattice 2024)\\
July 28th - August 3rd, 2024\\
University of Liverpool\\}

\begin{document}
\maketitle

\section{Introduction}
\vspace{-0.4cm}
The muon anomalous magnetic moment, $a_\mu = (g-2)_\mu /2$ where $g$ is the gyromagnetic ratio, has historically attracted much attention as a potential place to discover new physics~\cite{Aoyama:2020ynm}. The apparent tension between theory and experiment that for long had persisted is now questioned, given recent years' progress in e.g.~lattice quantum chromodynamics (QCD)~\cite{Borsanyi:2020mff,Boccaletti:2024guq}. % and experimental measurements of $e^+ e^-$ data needed for the traditional dispersive predictions of $a_\mu$~\cite{CMD-3:2023rfe}. 
It is important to understand the origin of the issue, and also for additional lattice QCD calculations to predict the hadronic vacuum polarisation (HVP) contribution to $a_\mu$ including isospin-breaking effects\footnote{There were several talks about this at the conference.}. 

Isospin-breaking effects arising from non-degenerate light-quark masses and electromagnetism typically enter as per cent level corrections, meaning that they have to be included for precision goals beyond that. %Examples beyond the hadronic vacuum polarisation are the leptonic decay rates in flavour physics, whose isospin-breaking corrections have been non-perturbatively calculated in~\cite{DiCarlo:2019thl,Boyle:2022lsi}.  
%These effects can be added perturbatively to lattice simulations, by expanding the underlying action around the isospin-symmetric point. While quark-mass differences are straightforward but computationally demanding to introduce, 
The long-range nature of the electromagnetic effects forbids charged states in finite-volume spacetimes with periodic boundary conditions~\cite{Hayakawa:2008an}. This underlying problem is related to Gauss' law, the absence of a mass gap in quantum electrodynamics (QED) and photon zero-momentum modes~\cite{Hayakawa:2008an}. However, the issue can be circumvented by defining finite-volume prescriptions for QED, such as $\qedl$ and infrared-improvement schemes~\cite{Hayakawa:2008an,Davoudi:2018qpl,Hermansson-Truedsson:2023krp,DiCarlo:2024lue}, $\qedc$~\cite{Lucini:2015hfa}, $\qedm$~\cite{Endres:2015gda} and $\textrm{QED}_\infty$~\cite{Asmussen:2016lse,Feng:2018qpx}. 

For $\qedl$ and $\qedc$, there generally are finite-volume effects (FVEs) scaling as inverse powers of the spatial volume extent, $1/L$. To extract physical predictions from lattice data, it is often useful to analytically subtract FVEs determined using effective field theory techniques~\cite{BMW:2014pzb,Lubicz:2016xro,Davoudi:2018qpl,Bijnens:2019ejw,DiCarlo:2021apt,Hermansson-Truedsson:2023krp,DiCarlo:2024lue} and fit the residual volume-dependence from the numerical data. For the HVP, it was in Ref.~\cite{Bijnens:2019ejw} shown using pointlike scalar $\qedl$ that the leading effects start at order $1/L^3$, as expected from the neutrality of the current~\cite{Giusti:2017jof}. Moreover, Ref.~\cite{Bijnens:2019ejw} argued from the analytical properties of the hadronic light-by-light tensor~\cite{Colangelo:2015ama,Colangelo:2017fiz} that the internal structure of the pion does not alter the cancellation at order $1/L^2$. In this talk, we take the first steps to determine the leading structure-dependent FVEs for the HVP in $\qedl$, arising at order $1/L^3$ in the large-volume expansion. 

\vspace{-0.3cm}
\section{Structure dependence in finite-volume effects}
\vspace{-0.4cm}
Let us consider an observable $\mathcal{O}(L)$ in lattice QCD+QED. We will neglect finite-time effects and only consider continuous Euclidean spacetime geometries $\mathbb{R}\times L^3$. %where the three spatial directions are periodic. %Here we will exemplify with $\qedl$, but this could be generalised to other prescriptions~\cite{Hermansson-Truedsson:2023krp,DiCarlo:2024lue}. 
The FVEs are given by $\Delta \mathcal{O}(L) = \mathcal{O}(L)-\mathcal{O}(\infty)$. The volume dependence can be obtained from a skeleton expansion of the underlying correlation function, which will generate a set of Feynman diagrams with one-particle irreducible vertex functions that depend on physical particle properties such as masses, charges and structure in terms of form factors~\cite{DiCarlo:2021apt}.%The structure dependence is then introduced by doing form-factor decompositions of the irreducible vertex functions~\cite{DiCarlo:2021apt}, which allows to calculate $\Delta \mathcal{O} (L)$.
\footnote{For an alternative but equivalent procedure, see the talk~\cite{DiCarloLattice:2024} at this conference. }
%introducing a generic effective field theory containing the interacting degrees of freedom, e.g.~pions and photons, allowing to calculate $\Delta \mathcal{O} (L)$. 
%The FVEs $\Delta \mathcal{O}(L)$ will depend only on physical particle properties such as masses, charges and structure in terms of form factors~\cite{DiCarlo:2021apt}. 
At leading order in QED, i.e.~order $e^2$, diagrams with virtual QED corrections will contain one photon, and consequently $\Delta \mathcal{O}(L)$ can be written
\begin{align}\label{eq:sumintdiffO}
	\Delta \mathcal{O}(L) = \left\{ \frac{1}{L^3}\sum _{\mathbf{k} \neq \mathbf{0}} - \int \frac{d^3 \mathbf{k}}{(2\pi)^3}\right\}
	\int \frac{dk_0}{2\pi } \frac{f_{\mathcal{O}}(k_0,\mathbf{k})}{k_0^2 + \mathbf{k}^2} \, ,
\end{align}
where the photon momentum $k = (k_0, \mathbf{k})$ has been introduced. Our choice of $\qedl$ here is manifested in terms of the absence of $\mathbf{k} = \mathbf{0}$ in the sum. The function $f_{\mathcal{O}}(k_0,\mathbf{k})$ depends on the observable $\mathcal{O}$ and in particular the physical properties of the particles in the process. One should further note that $f_{\mathcal{O}}(k_0,\mathbf{k})$ can contain analytical structure in $k_0$ as well, in particular poles from intermediate particles propagating and branch-cuts from form factors. Assuming that there are no infrared divergences in $\mathcal{O}$ or external spatial momenta, in a large-$L$ expansion the quantity $\Delta \mathcal{O}(L) $ takes the form~\cite{DiCarlo:2021apt,Hermansson-Truedsson:2023krp,DiCarlo:2024lue}
\begin{align}
	\Delta \mathcal{O}(L)  = \frac{c_{2}\, A_{2}}{m_\pi L}
	+ \frac{c_{1}\, A_{1}}{(m_\pi L)^2}
	+ \frac{c_{0}\, A_{0}}{(m_\pi L)^3}
	+\ldots \, .
\end{align}
Here exponentially suppressed terms $e^{-m_\pi L}$ as well as power-suppressed effects of order $1/(m_\pi L)^4$ have been neglected. The $c_j$ in the numerators are dimensionless finite-volume coefficients defined e.g.~in Refs.~\cite{Davoudi:2018qpl,DiCarlo:2021apt}.
%\begin{align}
%	c_j = \left\{ \sum _{\vec{n}\neq \mathbf{0}}-\int d^3 \vec{n}\right\} \frac{1}{|\vec{n}|^j} \, . 
%\end{align}
The $A_{j}$ contain the physics, in particular structure. It was observed in Ref.~\cite{DiCarlo:2021apt} that $A_0$ contains structure-dependent contributions associated to branch-cuts in the underlying correlation function. These cuts are difficult to estimate, which means that it is challenging to subtract $\Delta \mathcal{O}(L)$ in analyses of lattice data beyond order $1/(m_\pi L)^2$, see e.g.~Refs.~\cite{DiCarlo:2021apt,Boyle:2022lsi,Hermansson-Truedsson:2023krp,DiCarlo:2024lue}. %\footnote{We stress that it is important to, in addition to subtracting analytical FVEs, perform fits over lattice data at several physical volumes to pin down the residual volume dependence.}. 
As will be discussed below, for the HVP the expansion starts at order $1/(m_\pi L)^3$, meaning that unless one pins down the structure-dependence and cuts, the full leading correction cannot be subtracted. 

\vspace{-0.3cm}
\section{Hadronic vacuum polarisation}
\begin{figure}[t!]
	\centering
	\includegraphics[width=0.7\linewidth]{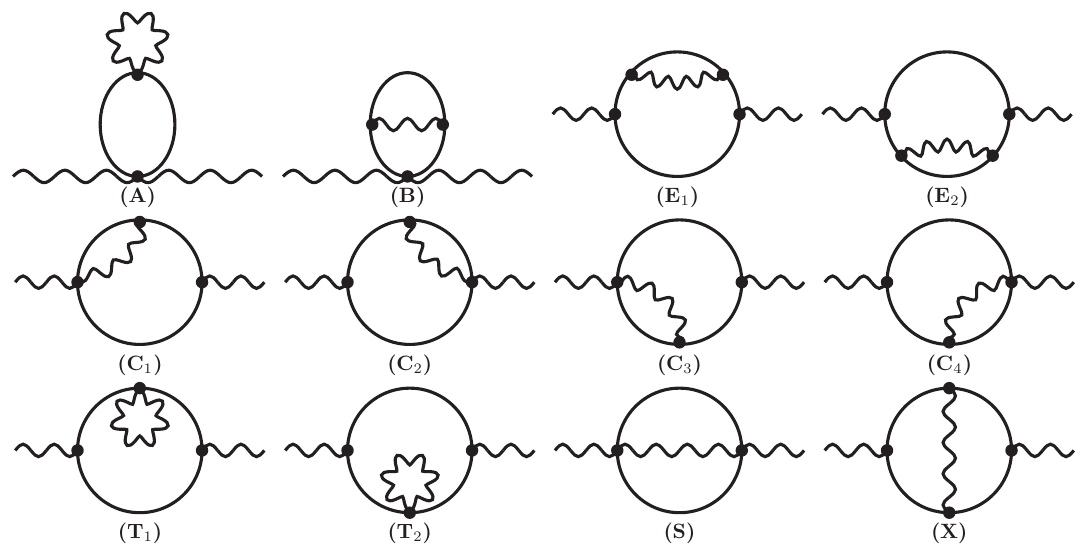}
	\caption{The 12 diagrams contributing at order $e^2$. %In practice only 7 topologies are independent, $A$, $B$, $E$, $C$, $T$, $S$ and $X$. 
	}
	\label{fig:DiagramNLOFig}
\end{figure}

\vspace{-0.4cm}
The HVP is defined as the vector-vector 2-point function
\begin{align}\label{eq:hvpcorrfun}
	\Pi _{\mu \nu}(q) = \int d^{4}x \, e^{iq\cdot x} \langle \left.  0\right| T \big[  J_{\mu} (x) J_{\nu }^{ \dagger } (0)\big]  \left| 0 \right.  \rangle \, ,
\end{align}
where $J_{\mu} (x)$ is the electromagnetic current and $q = (q_0, \vec{q} )$ is an external momentum. In the following, we will consider the kinematical setting $q = (q_0, \mathbf{0})$ with $q^2 >0$ in Euclidean space.  From the Ward-Takahashi identity $q_{\mu} \Pi _{\mu \nu} = 0$ it follows that $\Pi _{\mu \nu}(q^{2})$ is transverse, namely $\Pi _{\mu \nu}(q^{2}) = \left( q_{\mu}q_{\nu} - q^{2}\delta _{\mu \nu }\right) {\color{black}\Pi (q^{2})}$. We are interested in the subtracted and hence ultraviolet finite quantity
\begin{align}\label{eq:qhatdef}
		{ \color{black} \hat{ \Pi } (q^{2})} & = {\color{black}\Pi (q^{2})} -{\color{black}\Pi (0)} = \frac{1}{3q_{0}^{2}}\sum _{j=1}^{3}\Big( \Pi _{jj}(q^{2}) -\Pi _{jj}(0) \Big) \,.
\end{align}
The corresponding FVEs are then given by\footnote{Although this choice differs from the time-momentum representation approach typically employed in lattice calculations of the HVP~\cite{Aoyama:2020ynm}, the FVEs can be used when integrating $\Delta \hat{ \Pi } (q^{2},L )$ with the appropriate kernel to get the contribution to $a_\mu$. } $\Delta 	{  \hat{ \Pi } (q^{2},L)}  = 	{  \hat{ \Pi } (q^{2},L)}-	{ \hat{ \Pi } (q^{2},\infty)}$.
%\begin{align}
%	\Delta 	{  \hat{ \Pi } (q^{2},L)}  & = 	{  \hat{ \Pi } (q^{2},L)}-	{ \hat{ \Pi } (q^{2},\infty)}  \, .
%\end{align}
At order $e^2$, there are 12 diagrammatic topologies contributing, here shown in Fig.~\ref{fig:DiagramNLOFig}, but in practice only 7 are independent ($A$, $B$, $E$, $C$, $T$, $S$ and $X$). Separated into diagrams, we have 
\begin{align}
	\Delta 	{  \hat{ \Pi } (q^{2},L)}  & = 		
	\Delta 	{  \hat{ \Pi }_{A} (q^{2},L)}
	+	\Delta 	{  \hat{ \Pi }_{B} (q^{2},L)}
	+	2\, \Delta 	{  \hat{ \Pi }_{E} (q^{2},L)}
	+4\, 	\Delta 	{  \hat{ \Pi }_{C} (q^{2},L)}
	+ 2\, 	\Delta 	{  \hat{ \Pi }_{T} (q^{2},L)}
	\nonumber \\
	& 
	+ \Delta 	{  \hat{ \Pi }_{S} (q^{2},L)}
	+	\Delta 	{  \hat{ \Pi }_{X} (q^{2},L)}
	\, .
\end{align}
The vertices in the Feynman diagrams of Fig.~\ref{fig:DiagramNLOFig} correspond to the structure-dependent irreducible vertex functions obtained from a skeleton expansion of the correlation function in Eq.~(\ref{eq:hvpcorrfun}). 
These all contain two pions and in addition one or two photons, which we respectively denote $\Gamma _{\mu} (p,k)$ and $\Gamma _{\mu \nu} (p,k,q)$ for incoming pion momentum $p$, incoming photon momentum $k$ and outgoing photon momentum $q$. The form-factor decompositions of the vertex functions are known from virtual Compton scattering~\cite{Fearing:1996gs}, and given by
	\begin{align}\label{eq:ffdec}
	\Gamma _{\mu \nu} (p,k,q) 
& =
2\delta _{\mu \nu} [1-F(k^2)-F(q^2)]
-2\, k_\mu k_\nu \, \frac{1-F(k^2)}{k^2}
%\nonumber \\
%& 
-2\, q_\mu q_\nu \, \frac{1-F(q^2)}{q^2}
+ {\color{black}\Gamma _{\mu \nu}^{\textrm{T}} (p,k,q) } \, .
\nonumber \\
	\Gamma _{\mu} (p,k)  & = (2p+k)_\mu \, F(k^2) + k_\mu \frac{(p+k)^2-p^2}{k^2}\, [1-F(k^2)] \, ,
\end{align}
Here $F(k^2)$ is the electromagnetic form factor of the pion, with $F(0) = 1$ being the charge and $F'(0) = \langle r^2_{\pi}\rangle /6  $ proportional to the charge radius. The function $\Gamma _{\mu \nu}^{\textrm{T}} (p,k,q)$ is transverse with respect to the photon momenta, i.e.~$k_\mu\Gamma _{\mu \nu}^{\textrm{T}} (p,k,q) = -q_\nu \Gamma _{\mu \nu}^{\textrm{T}} (p,k,q) =  0$, and is purely structure dependent. There are 5 form factors in this transverse quantity~\cite{Fearing:1996gs} $G_1$, $G_2$, ..., $G_5$, each being a function of $k\cdot q$, $k^2+q^2$, $k^2-q^2$, $(k+q)\cdot (2p+k-q)$. Physically these are e.g.~related to the pion electromagnetic polarisabilities $\bar{\alpha}$ and $\bar{\beta}$. For brevity, we refrain from writing down the whole expression which can be obtained from section IV of Ref.~\cite{Fearing:1996gs}.\footnote{Eq.~(\ref{eq:ffdec}) only depends on on-shell information, which is a choice since $\Delta \hat{\Pi} (q^2,L)$ only can depend on physical quantities. Consequently, it would be independent of any off-shellness in the form factors~\cite{BMW:2014pzb,Lubicz:2016xro,DiCarlo:2021apt}, %The similar situation occurs for the pseudoscalar masses~\cite{BMW:2014pzb,Lubicz:2016xro,DiCarlo:2021apt} and leptonic decay rates~\cite{Lubicz:2016xro,DiCarlo:2021apt}, which 
	as can be understood from the equivalence between the skeleton expansion and on-shell approaches such as  in the talk~\cite{DiCarloLattice:2024}.} The pointlike scalar QED calculation in Ref.~\cite{Bijnens:2019ejw} can be obtained from these vertex functions through the limit $F(k^2) = F(q^2) =  1$ and $\Gamma _{\mu \nu}^{\textrm{T}} (p,k,q) = 0$. 

Having thus defined the structure-dependent vertex functions, we may proceed to evaluate the respective contributions $\Delta \hat{\Pi } _{U}(q^2,L)$. Since these diagrams have two loops and the pions are also in finite-volume, $\Delta \hat{\Pi } _{U}(q^2,L)$ takes the form
\vspace{-0.2cm}
\begin{align}
	& \Delta \hat{\Pi } _{U} = \left( \frac{1}{L^3}\left. \sum _{\mathbf{k}\neq \mathbf{0}}\right.  \, \frac{1}{L^3}\sum _{\boldsymbol{\ell}}- \int \frac{d^{3} \mathbf{k}}{(2\pi)^{3}}\int \frac{d^{3}\boldsymbol{\ell}}{(2\pi)^{3}} \right) \int \frac{dk_0}{2\pi } \frac{d\ell_0}{2\pi }\, \hat{\pi }_{U}\left( q^{2}_{0},k,\ell \right) \, . 
\end{align}
Here $\ell = (\ell _0 , \boldsymbol{\ell})$ is the pion momentum, and $\hat{\pi }_{U}\left( q^{2}_{0},k,\ell \right)$ is the integrand of Feynman diagram $U$ in Fig.~\ref{fig:DiagramNLOFig}. From the kinematical choice $q^2>0$, there are no kinematical singularities in the integrand associated to the pions, and we may therefore replace the sum over $\boldsymbol{\ell}$ with an integral, which is valid up to corrections exponentially suppressed with the volume~\cite{Bijnens:2019ejw}. We thus have the simple sum-integral difference over the photon momentum as in Eq.~(\ref{eq:sumintdiffO}), i.e.
\begin{align}\label{eq:fvesphoton}
	& \Delta \hat{\Pi } _{U}(q^2,L) = \left( \frac{1}{L^3}\left. \sum _{\mathbf{k}\neq \mathbf{0}}\right.  - \int \frac{d^{3} \mathbf{k}}{(2\pi)^{3}} \right) \int \frac{d^3 \boldsymbol{\ell}}{(2\pi)^3}\,  \int \frac{dk_0}{2\pi } \frac{d\ell_0}{2\pi }\, \hat{\pi }_{U}\left( q^{2}_{0},k,\ell \right) +\ldots \, . 
\end{align}
As examples of integrands, we have diagrams $E$ and $X$,
\begin{align}\label{eq:fveshvp}
	\hat{\pi }_{E}\left( q^{2}_{0},k,\ell \right) 
	& = 
	\frac{1}{3 q_0^2  }
	\left\{
	\frac{\Gamma _{j}(\ell -q, q)\, \Gamma _{\mu}(\ell , k) \Gamma _{\mu}(k+\ell , -k) \,  \Gamma _{j}(\ell , -q)}{k^2\, [\ell^2 + m_\pi^2]^2\, [(k+\ell )^2+m_\pi^2]\, [(q-\ell)^2+m_\pi^2]} 
	\right\} _{q}
	\, ,
	\\
		\hat{\pi }_{X}\left( q^{2}_{0},k,\ell \right) 
	& = 
	\frac{1}{3 q_0^2  } \left\{
	\frac{\Gamma _{j}(\ell -q, q)\, \Gamma _{\mu}(\ell , k) \Gamma _{\mu}(k+\ell , -q) \,  \Gamma _{j}(\ell+k-q , -k)}{ k^2\, [\ell^2 + m_\pi^2]\, [(k+\ell )^2+m_\pi^2]\, [(q-\ell)^2+m_\pi^2]\, [(q-k-\ell)^2+m_\pi^2]}
	\right\} _{q}
	\, . 
\end{align}
Here the subscript $q$ indicates that whatever is in the curly brackets has to have a subtraction at $q^2 = 0$ in accordance with Eq.~(\ref{eq:qhatdef}). 

In the pointlike scalar QED limit, one should find~\cite{Bijnens:2019ejw}
\begin{align}\label{eq:hvppt}
	\Delta \hat{\Pi} (q^2) \stackrel{\textrm{point}}{=}
	\frac{c_0}{(m_{\pi} L)^3}
	\Bigg(
	\frac{16}{3}\, \Omega_{0,3}
	+ \frac{5}{3}\, \Omega_{2,2}
	- \frac{40}{9}\, \Omega_{2,3}
	+ \frac{3}{8}\, \Omega_{4,1}
	- \frac{7}{6}\, \Omega_{4,2}
	- \frac{8}{9}\, \Omega_{4,3}
	\Bigg)\, .
\end{align}
where the integrals $\Omega _{i,j} = \Omega _{i,j}(q_0^2/m_\pi^2)$ are defined in Ref.~\cite{Bijnens:2019ejw}. Diagram by diagram there are also $1/(m_\pi L)^2$ terms, but these cancel in the full sum due to the neutrality of the currents in the HVP~\cite{Giusti:2017jof,Bijnens:2019ejw}. As was also argued in Refs.~\cite{Giusti:2017jof,Bijnens:2019ejw}, even in the structure-dependent case we should see that $\Delta \hat{\Pi} (q^2)$ starts at order $1/(m_\pi L)^3$. We briefly note that in  $\qedr$~\cite{Hermansson-Truedsson:2023krp,DiCarlo:2024lue} and $\qedc$~\cite{Lucini:2015hfa} the leading effects start at order $1/(m_\pi L)^4$, since the equivalent of $c_{0}$ there is zero.  

\vspace{-0.4cm}
\section{Towards an evaluation of the finite-volume effects}
\vspace{-0.3cm}
Next we discuss the prospects of evaluating the leading FVEs including structure dependence in $\Delta \hat{\Pi}(q^2,L)$, and report on some preliminary findings. It should be noted that the $k_0 $ and $\ell _0$ integrals in Eq.~(\ref{eq:fvesphoton}) pick up all the analytical structure in the integrand $\hat{\pi }_{U}\left( q^{2}_{0},k,\ell \right)$, i.e.~both pole and branch-cuts. We may then separate $	\Delta \hat{\Pi} (q^2,L) $ into pure pole contributions and a remainder,
\begin{align}
	\Delta \hat{\Pi} (q^2,L) = \Delta _{\textrm{poles}}\hat{\Pi} (q^2,L) + \Delta _{\textrm{rem}}\hat{\Pi} (q^2,L) \, .
\end{align}
The pole contributions can be directly evaluated from the singularities in $k_0$ and $\ell _0$ from the propagators in the integrands $ \hat{\pi }_{U}\left( q^{2}_{0},k,\ell \right) $. Doing this and a large-volume expansion one then obtains e.g., for the sum of diagrams $E$ and $X$,
\begin{align}
	& \Delta _{\textrm{poles}}\hat{\Pi}_{E+X} (q^2,L) 
	= 
	\frac{c_1}{24 \pi z\,  
		(m_\pi L)^2 }\,
	\Bigg\{
	4 \Big[ 4 z^2 \, \Omega _{1,3}-7 z^2\,  \Omega _{3,3}+3 z^2 \, \Omega _{5,3}-96
	z \, \Omega _{1,3}
\nonumber 	\\
	& \qquad 
	+40 z \, \Omega _{3,3}+8 (7 z-66)\,  \Omega _{-1,3}+288
	\Omega _{-3,3}+240\,  \Omega _{1,3}\Big] {\color{black}F (q_0^2 )^2}
	\nonumber \\
& \qquad 
	-3 \Big[
	6 z^3 \, \Omega _{3,3}
	-11 z^3 \, \Omega _{5,3}
	+5 z^3 \, \Omega _{7,3}
	+72 z^2 \, \Omega _{1,3}
	-132 z^2 \, \Omega _{3,3}
	+60 z^2 \, \Omega _{5,3}
		\nonumber \\
	& \qquad 
	-528 z \, \Omega _{1,3}
	+240 z \, \Omega _{3,3}
	+32 (9 z-22) \, \Omega _{-1,3}
	+384 \, \Omega _{-3,3}
	+320 \, \Omega _{1,3}
	\Big]
	\Bigg\}
	+ \frac{c_{0}\, \mathcal{C}^{\textrm{poles}}_{E+X}}{(m_\pi L)^3} 
	%+ \mathcal{O} \left[\frac{1}{(m_\pi L)^3}, e^{-m_\pi L}\right] 
	\, .
\end{align}
Here $z = q_0^2 /m_\pi ^2$. We have left out an explicit expression of the structure dependent $\mathcal{C}^{\textrm{poles}}_{E+X}$ due to its length. By setting $F(q_0^2)=1$ and $F'(0) = 0$ for the $1/(m_\pi L)^3$ contribution we regain the pointlike scalar QED result from Ref.~\cite{Bijnens:2019ejw}. One may proceed in this way to evaluate the full $\Delta _{\textrm{poles}}\hat{\Pi} (q^2,L)$ through order $1/(m_\pi L)^3$, which in the pointlike limit should give back Eq.~(\ref{eq:hvppt}). 

The issue to evaluate $ \Delta _{\textrm{rem}}\hat{\Pi} (q^2,L)$ remains, and is crucial since it is structure-dependent as well and can give cancellations with $ \Delta _{\textrm{poles}}\hat{\Pi} (q^2,L)$, thus altering the size of  $\Delta \hat{\Pi} (q^2,L)$. To obtain $ \Delta _{\textrm{rem}}\hat{\Pi} (q^2,L)$ we propose to exploit the connection between the HVP and the hadronic light-by-light tensor $\Pi _{\mu \nu \rho \sigma} (q_1,q_2,q_3,q_4)$ in forward kinematics,
\begin{align}
	 \hat{\Pi} (q^2) = \frac{1}{3q_0^2} \int \frac{d^4 k}{(2\pi)^4} \, \frac{\hat{\Pi}_{j j \mu \mu}(q,-q,k,-k)}{k^2} \, .
\end{align}
 In Ref.~\cite{Biloshytskyi:2022ets} the above relation is rewritten in terms of a dispersive sum rule for $\gamma ^\ast \gamma ^\ast$-fusion cross sections. Moreover, in Refs.~\cite{Colangelo:2015ama,Colangelo:2017fiz} it was proven that the two-pion contribution in the dispersive representation of the hadronic light-by-light is in one-to-one correspondence with the Feynman diagram approach involving two pions (equivalent to the approach here). The proposed way forward is therefore to evaluate $\Delta _{\textrm{poles}}\hat{\Pi} (q^2,L) $ using the form-factor decompositions in Eq.~(\ref{eq:ffdec}), and $\Delta _{\textrm{rem}}\hat{\Pi} (q^2,L) $ by connecting it to dispersion theory for the hadronic light-by-light. 

\vspace{-0.4cm}
\section{Conclusions}
\vspace{-0.3cm}
We have discussed the prospects of evaluating the leading FVEs to the HVP in $\qedl$. This is an extension of Ref.~\cite{Bijnens:2019ejw} to include also structure dependence through form factors as in Ref.~\cite{DiCarlo:2021apt}. The work is relevant for evaulation of the leading isospin-breaking corrections to the HVP contribution to the muon anomalous magnetic moment.  
%$\qedr$~\cite{Hermansson-Truedsson:2023krp,DiCarlo:2024lue} and $\qedc$~\cite{Lucini:2015hfa}

\vspace{-0.4cm}
\section*{Acknowledgements}
\vspace{-0.4cm}
N.~H.-T.~ wishes to thank M.~Di Carlo, M.~T.~Hansen and A.~Portelli for useful discussions and collaboration over the years, M.~Bruno and C.~Lehner for interesting discussions about possible numerical validations of this work, as well as J.~Bijnens, J.~Harrison, T.~Janowski, A~J\"{u}ttner and A.~Portelli for the collaboration in Ref.~\cite{Bijnens:2019ejw} forming the basis for this work. 
N.~H.-T.~is funded by the UKRI, Engineering and Physical Sciences Research Council, grant number EP/X021971/1. 

%\newpage
%Bibliography
\bibliographystyle{JHEP}
\bibliography{refs}

\end{document}